\documentclass[12pt,a4paper]{article}
\usepackage[utf8]{inputenc}
\usepackage[T1]{fontenc}
\usepackage{alphalph}
\usepackage{setspace}
\usepackage{textcomp}
\usepackage{amsmath}
\usepackage{amsfonts}
\usepackage{amssymb}
\usepackage[]{graphicx}
\usepackage{authblk}
\alph{subsection}

\begin{document}
\date {{\small {\today}}}

\renewcommand\Authfont{\fontsize{14}{14.4}\selectfont}
\renewcommand\Affilfont{\fontsize{9}{10.8}\selectfont}
\renewcommand\Authand{ \textbf{.} }

\author[1,2]{Hamid Saleem}
\author[3]{Usman Saeed}
\affil[1]{Department of Space Science, Institute of Space Technology (IST), 1-Islamabad Highway near CDA toll plaza, Islamabad, Pakistan.}
\affil[2]{Pakistan Academy of Sciences, 3-Constitution Avenue, G-5/2, Islamabad, Pakistan.}
\affil[3]{National Centre for Physics,
QAU Campus, Shahdra Valley Road,
P.O. Box No. 2141, Islamabad - 44000,
Pakistan.}

\title{\textbf{Extraordinary Drift Wave \\
in Space and Cylindrically Bounded\\
Heavier Ion Plasmas}}
\maketitle

\begin{abstract}
It is pointed out that electrostatic and electromagnetic waves propagating in perpendicular direction to the density gradient $\nabla n_0=\hat{x} |\frac{dn_0}{dx}|$ and external magnetic field $\vec{B}_0=B_0 \hat{z}$ in hybrid frequency range $\Omega_i \ll \omega \ll \Omega_e$ (where $\Omega_j=\frac{eB_0}{m_j c}$ and $j=e,i$) have several applications in space and laboratory plasmas. The electron plasma waves are important for the future experiments on heavier ion plasmas as well as the results are applicable to the experiments performed to create pure pair ion  fullerene plasmas. The electrostatic modes are shown to exist in upper F-region of terrestrial ionosphere and electromagnetic waves are applied to cylindrically bounded heavier ion plasmas. 
\end{abstract}
\doublespacing
\section {Introduction}

The electrostatic drift type waves in hybrid frequency range
($\Omega_i<\omega<\Omega_e$), where $\Omega_j=\frac{eB_0}{m_j c}$ is
the gyro frequency of j-th species and $j=(e,i)$ have been
investigated in the recent past (Eliasson et al. 2006; Saleem and Eliasson 2011) in the light of future plans
to perform experiments with heavier ions. The study of heavier ion
plasmas is also important due to their relevance to experiments
performed to produce pair ion fullerene plasmas (Oohara and Hatakeyama 2003; Oohara et al. 2005; Oohara eta al. 2009) which have
generated controversy among theoreticians (Saleem 2006; Verheest 2006; Kono eta al. 2014).

In heavier ion plasmas the gap between cyclotron frequencies of ions
$\Omega_i$ and electrons $\Omega_e$ is large and hence some waves
may appear in the frequency range $\Omega_i \ll\omega < \Omega_e$ and
for these waves the ion dynamics can be ignored. However, for the
space plasmas of hydrogen or fusion plasmas of deuterium (D)-tritium
(T), the role of ions is important in the hybrid frequency waves due
to small ratio of masses $\frac{m_e}{m_i}$.

More than a decade ago (Huba 1991; Huba eta al. 1995), a purely transverse wave, the
magnetic drift wave (MDW), was proposed to exist in inhomogeneous
density electron plasmas with linear dispersion relation
\begin{equation}
\omega= k_y v_A(\frac{c}{\Omega_{i}
L_n})=\lambda_e^2k_y^2(\frac{\kappa_n}{k_y} \Omega_e)
\end{equation}
where  $\lambda_e=\frac{c}{\omega_{pe}}$ is the electron skin
depth, $\omega_{pe}=\sqrt{\frac{4 \pi n_0 e^2}{m_e}}$ is the
electron plasma frequency, $L_n=\frac{1}{\kappa_n}$ is the density
gradient scale length, and $v_A=(\frac{B_0^2}{4\pi m_i
n_0})^{\frac{1}{2}}$ is Alfven speed. The density gradient is
assumed to be in the positive x-direction $\nabla n_0={\hat{x}}\mid
\frac{dn_0}{dx}\mid $, ${\vec {B}}_0={\hat {z}}B_0$ and wave propagates in
y-direction perpendicular to both density gradient and external
magnetic field. The linear dispersion relation (1) was derived using
electron magnetohydrodynamics (EMHD) equations. In standard EMHD
model (Kingssep et al. 1990), the electrons are considered to be inertia-less
($m_e\rightarrow 0$) and ion dynamics is ignored in the limit
($m_i\rightarrow \infty$) for time scale $\tau$ such that
$\Omega_e^{-1}, \omega_{pe}^{-1} \ll \tau \ll \Omega_i^{-1},
\omega_{pi}^{-1}$ that is for frequency range $\Omega_i,\, \omega_{pi}\ll \omega
\ll \omega_{pe},\, \Omega_e$.

Recently (Saleem 2014), it has been proposed that one should use equation of
motion to determine the perpendicular electron velocity components
${\vec{v} }_e=(v_{ex},v_{ey})$ instead of using Ampere's law for ${\vec{v}}_e$ . If
electron velocity is determined appropriately using equations of
motion, then one notices that compressibility cannot be ignored in the MDW model. The
wave becomes partially transverse and partially longitudinal and in
the presence of density gradient its dispersion relation is modified
as

\begin{equation}
\omega=\frac{\lambda_e^2 k_y^2}{1+\lambda_e^2
k_y^2(1+\frac{\Omega_e^2}{\omega_{pe}^2})}(\frac{\kappa_n}{k_y}
\Omega_e)
\end{equation}

In heavier ion plasmas, this wave can play an important role and it
has not been investigated in such plasmas so far.
 In lighter element plasmas, the frequency window
$\Omega_i<\omega<\Omega_e$ is narrower and this wave couples with
the lower hybrid oscillations (Saleem 2014).
The electrostatic (ES) lower-hybrid wave (LHW) is important due to its
role in laboratory plasma heating and spacecraft observations
of terrestrial ionosphere and magnetosphere . In the presence of
density gradient, the LHW couples with the electrostatic electron
convection mode (ECM) (Yu et al. 1985). Both the LHW and ECM propagate obliquely
making a small angle with the external constant magnetic field satisfying $k_z\ll k_{\perp}$ and
$\Omega_i< \omega < \Omega_e$  while $k_z (k_{\perp})$
are parallel (perpendicular) components of the wave vector, and
$\omega$ is the wave frequency.

The pure transverse electromagnetic wave known as the ordinary \textbf{O}-
mode (Chen 1984) in plasmas couples with the transverse current and its
linear dispersion relation becomes $(\omega^2 = \omega_{pe}^2 +
c^2 k^2)$ . The extraordinary wave
(X-mode) is also an electron time scale wave and it is partially
transverse and partially longitudinal (Chen 1984). Both O and X modes are
high frequency and short wavelength waves and do not need ion
dynamics for their description, in general.

In this work, it is pointed out that the extraordinary (\textbf{X}-) mode in the low frequency limit becomes drift
type wave in an inhomogeneous density plasma which can couple with
the lower hybrid oscillations when the ion dynamics is taken into
account. This wave propagates in the direction perpendicular to the
external magnetic field and has frequency in the lower hybrid
frequency range $\Omega_i\ll\omega\ll\Omega_e$ .  The basic wave is described in cartesian geometry
and its electric field structure is found in a cylindrically bounded
plasma by solving the coupled Poisson equation and Ampere's law. In
heavier ion plasmas the ion dynamics may be ignored.
\\Here we mainly discuss the electromagnetic extraordinary drift wave in cylindrically bounded plasma for its application to future experiments on heavier ion and pair ion fullerene plasmas. Then the eigenvalue problem is also solved in a
cylindrically bounded plasma with Gaussian density distribution.


\section{Extraordinary Drift Wave}Let us consider the extra-ordinary wave with $\vec{B_0}=B_0\hat{z},\,\vec{k}=k_{y}\hat{y}$  and $ \vec{E}=\hat{x}E_{x}+\hat{y}E_{y} $ in an inhomogeneous plasma $\nabla n_{0}=-\hat{x}|\dfrac{dn_{0}}{dx}| $. The Poisson equation yields \begin{equation}
[\omega^{2}(1-\dfrac{\Omega^2_{e}}{\omega^2})-\omega^2_{pe}-\dfrac{\kappa_n}{k_y}\dfrac{\Omega_e}{\omega}\omega^2_{pe}]E_{1y}-\dot{\iota}\dfrac{\omega^2_{pe}}{\omega}(\Omega_e+\dfrac{\kappa_n}{k_y}\omega)E_{1x}=0
\end{equation}
and Maxwell's equation provides us with another relation between $ E_{1x} $ and $ E_{1y} $ as follows \begin{equation}
[(\omega^{2}-c^{2}k^{2})(1-\dfrac{\Omega^2_{e}}{\omega^2})-\omega^2_{pe}-\dfrac{\kappa_n}{k_y}\dfrac{\Omega_e}{\omega}\omega^2_{pe}]E_{1x}+\dot{\iota}\omega^2_{pe}(\dfrac{\Omega_e}{\omega}+\dfrac{\kappa_n}{k_y})E_{1y}=0
\end{equation}If we neglect $ \kappa_n $ terms, then (3) and (4) are similar to the set of equations (4-101) of (Chen 1984) which gives dispersion relation for X-mode as \begin{equation}
(\omega^2-\omega^2_H)[\omega^2-\omega^2_H-c^{2}k^{2}(1-\dfrac{\Omega^2_e}{\omega^2})]=(\dfrac{\omega^2_{pe}\Omega_e}{\omega})^2
\end{equation}where $ \omega^2_H=(\omega^2_{pe}+\Omega^2_e) $ is the upper hybrid frequency wave. The \textbf{X}-mode is well studied in literature but the low frequency regime of this wave in the presence of density gradient has not been investigated. In lighter ion plasmas, the wave frequency becomes closer to $ \Omega_i $ or $(\Omega_e\Omega_i)^{1/2}$ in the limit $\omega^2\ll\omega^2_{pe},\Omega^2_{e}$ , therefore the role of ion dynamics must be included. In this case Eq. (2) becomes, \begin{equation}
\{1+\lambda^2_e k^2_e(1+\dfrac{\Omega^2_e}{\omega^2_{pe}})\}\omega^2+\lambda^2_e k^2_e(\dfrac{\kappa_n}{k_y}\Omega_e)-\lambda^2_e k^2_e(\Omega_e \Omega_i)=0
\end{equation} Long ago (Yu et al. 1985), the electrostatic electron convection mode was investigated which has the dispersion relation\begin{equation}
\omega=\omega_{\upsilon}[1+\dfrac{m_i}{m_e}\dfrac{k^2_z}{k^2_y}]
\end{equation}where \begin{equation}
\omega_\upsilon=(k_yL_n)\Omega_i
\end{equation} for $k_z\ll k_y$.\\The coupled dispersion relation of electron convection mode and lower hybrid wave was presented in equation (5) of Ref. \cite{13} as, \begin{equation}
1-\dfrac{\omega^2_{pi}}{\omega^2}+\dfrac{\omega^2_{pe}}{k^2_y}[\dfrac{k_y}{L_n \omega \Omega_e}-\dfrac{k^2_z}{\omega^2}+\dfrac{k^2_y}{\Omega^2_e}]=0
\end{equation}Yu et al. (1985) discuss the obliquely propagating two modes only; the lower hybrid wave and electron convection mode. \\It is pointed out here that the Eq. (9) also contains an electron mode which propagates purely in the perpendicular direction to the external magnetic field and density gradient.\\Let $k_z=0,\,k_x=0$, and put $\omega_{pi}=0$ (i.e. ions stationary), then Eq.(9) reduces to 
\begin{equation}
\omega=-\dfrac{1}{(1+\dfrac{\Omega^2_e}{\omega^2_{pe}})}(\dfrac{\kappa_n}{k_y}\Omega_e)
\end{equation} In heavier ion plasmas (like Barium or Fullerene), the above wave will be relevant. In lighter element plasmas, this will couple with the lower hybrid frequency and Eq.(10) will become,
\begin{equation}
(1+\dfrac{\Omega^2_e}{\omega^2_{pe}})\omega^2+\dfrac{\kappa_n}{k_y}\Omega_e-\Omega_i\Omega_e=0
\end{equation}
Both  dispersion relations (10) and (11) have been presented by Saleem (2014) but it seems important to point out that these simple dispersion relations were present in earlier literature implicitly.
\section{Cylindrically Bounded Plasmas}
The waves discussed in previous section have different regimes of application. The electrostatic modes have good applications both in space and laboratory plasmas, while the electromagnetic modes are more relevant to plasmas with $ 1\lesssim \lambda^2_e k^2_y $ and this limit is generally satisfied in laboratory plasmas.\\The heavier ion fullerene plasmas have been produced in a cylinder of length $ l=90 $ cm and diameter $ R_0=3 $ cm (Oohara et al 2005). The planned Barium plasma experiments will probably be performed in a cylindrical geometry (Q Machine).\\Therefore, in this section, we present the form of local modes in cylindrical geometry and plasma is assumed to have Gaussian density profile,
\begin{equation}
n_0(r)=N_0e^{-r^2/R^2_0}
\end{equation} where $R_0$ is the cylinder radius and $N_0$ is the density at the axis. The external magnetic field is along the $z$-axis and waves propagate in $\theta$-direction. The external magnetic field is along $z$-axis, the perturbed electric and magnetic fields are, respectively, $\vec{E}=\hat{r}E_r+\hat{\theta}E_{\theta}$ and $\vec{B}=\hat{z}B$. The perturbations are assumed to have the form $\psi=\psi(r)e^{\dot{\iota}(m\theta-\omega t)}$ where $m$ is the azimuthal wave number and $k_\theta=\dfrac{m}{r}$.\\Poisson equation takes the form, 
\begin{multline}
(L^2_{es}-\dfrac{\omega\Omega_e}{m})E_{\theta}-(\dfrac{\omega\Omega_e r}{m})\partial_{r}E_{\theta}\\=\{-\omega\Omega_e+\dfrac{L^2_0}{m}+(\omega^2-\Omega_i\Omega_e)\dfrac{r\kappa_n}{m}\dot{\iota}E_r+L^2_0 \dfrac{r}{m}\dot{\iota}\partial_{r}E_r\}
\end{multline}

where \begin{align*}
L^2_{es}&=(L^2_0-\dfrac{r\kappa_n}{k_\theta}\Omega_e\omega),\\L^2_0&=(1+\dfrac{\Omega^2_e}{\omega^2_{pe}})\omega^2-\Omega_i\Omega_e\,.
\end{align*} In the limit $\omega \ll \Omega_e\,,\,\omega_{pe}$ , Ampere's law gives,
\begin{equation}
E_r=-\dot{\iota}\dfrac{r^2}{\lambda^2_e m^2}\{(\dfrac{\omega}{\Omega_e}+\dfrac{\lambda^2_e}{r^2} m)E_{\theta}+m\dfrac{\lambda^2_e}{r}\partial_r E_{\theta}\}
\end{equation}
\subsection{Local modes in cylinder}
The density profile (12) gives in cylindrical geometry $\kappa_n=-\dfrac{2r}{R_0}$, therefore the local wave dispersion relation takes the form,
\begin{equation}
\{1+(1+\dfrac{\Omega^2_e}{\omega^2_{pe}})\dfrac{\lambda^2_e}{r^2}m^2\}\omega^2+(2m\dfrac{\lambda^2_e}{R^2_0})\Omega_e\omega-\Omega_i \Omega_e \dfrac{\lambda^2_e}{r^2} m^2 =0
\end{equation} which in a heavier ion plasma with static ions becomes
\begin{equation}
\omega=- \dfrac{(2m\dfrac{\lambda^2_e}{R^2_0})\Omega_e}{\{1+(1+\dfrac{\Omega^2_e}{\omega^2_{pe}})\dfrac{\lambda^2_e}{r^2}m^2\}}
\end{equation} In electrostatic limit, Eq. (15) reduces to 
\begin{equation}
(1+\dfrac{\Omega^2_e}{\omega^2_{pe}})\omega^2+2\dfrac{r^2}{mR^2_0}\Omega_e\omega-\Omega_i \Omega_e=0
\end{equation} which in a heavier ion plasma gives,
\begin{equation}
\omega=-\dfrac{(2m\dfrac{r^2}{R^2_0})\Omega_e}{(1+\dfrac{\Omega^2_e}{\omega^2_{pe}})}
\end{equation}
\subsection{Eigen Modes of X - drift wave }
Here we analyze the low frequency $X$-mode in a cylinder with small variation of amplitudes along radial direction assuming that the fields vanish at the boundary $r=R_0$. We use the conditions 
\begin{equation}
\kappa_n\partial_r E_{\theta} \ll E_{\theta}\,,\,|\kappa_n\partial_r E_{\theta}| <|\dfrac{m^2}{r^2}\partial^2_r E_{\theta}|<m^2|E_{\theta}|
\end{equation} and consider the waves in the region $0\ll r \lesssim R_0$. \\ Then the Poisson equation and Ampere's law after simplification yield,\begin{equation}
r^2 d^2_r E_{\theta}+r d_r E_{\theta}+(\alpha^2r^2-m^2)E_{\theta}=0
\end{equation}where 
\begin{equation}
\alpha=\dfrac{(\dfrac{\omega^2}{\lambda^2_e}+2\dfrac{m}{R^2_0}\omega\Omega_e)}{\{\Omega_i \Omega_e-(1+\dfrac{\Omega^2_e}{\omega^2_{pe}})\omega^2\}}
\end{equation}The $r$-dependent general solution of Eq. (20) is 
\begin{equation}
E_{\theta}(r)=A_1J_m(\alpha r)+A_2Y_m(\alpha r)
\end{equation}where $J_m\,,\,Y_m$ are Bessel functions of order $m$ and $A_1,A_2$ are constant amplitudes.\\ At $r=0$, $Y_m$ blows up, therefore we put $A_2=0$ and the physical solution becomes,
\begin{equation}
E_{\theta}(r)=A_1J_m(\alpha r)
\end{equation}We choose the boundary condition $E_{\theta}=0$ at $r=R_0$, therefore $ J_m(\alpha R_0)=0 $ gives $k_{\alpha}=\alpha R_0$ where $ k_{\alpha} $ are the zeros of the Bessel function $J_m$.
\section{Applications} First the electrostatic waves are shown to exist in $ \mathrm{F} $-region of terrestrial ionosphere considering $n_0=7.5\times 10^3 \ {\mathrm{cm}}^{-3}$, $B_0=0.3$ Gauss in hydrogen plasma assuming the density gradient scale length $L_n=5 \times 10^4 $ cm\,=\,$500$ m (i.e., $\kappa_n=2\times 10^5 \, \mathrm{cm^{-1}}$) . This plasma has $\Omega_e=5.2 \times 10^6\, \mathrm{rads^{-1}}\,,\Omega_i=2.8\times 10^3\, {\mathrm{rads}}^{-1}\,,\,\lambda_e=6.1 \times 10^3\,\mathrm{cm}\,,\,\omega_{LH}=1.2 \times 10^5\, \mathrm{rads^{-1}}$ and we choose $k_y\,:\,(0.005-0.05)\,\mathrm{cm}$ so that $1\ll k^2_y \lambda^2_e$ holds and hence only electrostatic waves are investigated within local approximation $\kappa_n\ll k_y$ . In applications we shall use $\nabla n_0=\hat{x}|\frac{dn_0}{dx}|$ i.e. density gradient along $+$ ve $x$-axis and then $\omega$ of Eqs. (10) and (11) will be positive.

The Fig. (1a) shows $\omega$ vs $k_y$ corresponding to electron plasma dipersion relation Eq. (10). It is obvious from Fig. (1a) that $\Omega_i\lesssim \omega \ll \Omega_e$ and hence ion dynamics should be considered.
 

In Fig. (1b), the plot of $\omega$ vs $k_y$ of Eq. (11) is shown and $\Omega_i < \omega < \Omega_e$ is satisfied. The wavelength of the wave $\lambda_{\omega}=\frac{2\pi}{k_y}\,:\,(12-1)\, \mathrm{m}$ satisfies the condition $\lambda_{\omega}\ll L_n$ . For small $k_y\,:\,(10^{-3}-10^{-2}\,\mathrm{cm^{-1}})$ , the  wavelength becomes larger and Fig. (2a) and Fig. (2b) show, respectively, the plots $\omega$ vs $k_y$ of Eqs. (10) and (11). Corresponding to $k_y=10^{-3}$ , the wavelength becomes $\lambda_{\omega}\,=\,60\,\mathrm{m}$ and still $\lambda_{\omega}\ll L_n$ holds.
\\ The dispersion relations for electromagnetic waves in pure electron and electron-ion hydrogen plasmas are shown in Fig. (3a) and Fig. (3b), respectively using Cartesian geometry and JET parameters i.e. $n_0=2 \times 10^{14}\,\mathrm{cm^{-3}}\,,\,B_0=10^4 \,\mathrm{Gauss}$, and $\kappa_n=0.1 \,\mathrm{cm^{-1}}$ ($L_n=10\,\mathrm{cm}$). In this case, $\Omega_i=1.7 \times 10^{11}\,\mathrm{rads^{-1}}\,,\,\Omega_e=1.7 \times 10^{11}\,\mathrm{rads^{-1}}\,,\,\omega_{LH}=4.1 \times 10^9\,\mathrm{rads^{-1}}$ and $\lambda_e=0.0375\,\mathrm{cm}$. Since $\omega$ is closer to $\Omega_i$, therefore Fig. (3a) does not satisfy the static ions condition $\Omega_i\ll \omega$. The ion dynamics is needed and hence dispersion relation of  Eq.(6) is shown in Fig. (3b).\\ In case of JET parameters, the cylindrical coordinates are better to use. However, in this case, we take $R_0$ as average value of JET minor radius $R_0=1.67\,\mathrm{m}$. The Fig. (4a) shows the plot $\omega \,\mathrm{vs}\, r$ of Eq. (16) corresponding to $\mathrm{m}=500$ (dashed curve) and $\mathrm{m}=1000$ (solid curve). This EM wave is relevant for $0\ll r$ away from axis. Near axis, $\omega$ is closer to lower hybrid frequency, therefore ions should be considered to be dynamic as shown in Fig. (4b). Here we obtain $\omega<\Omega_i$. Therefore, the dispersion relation of Eq. (16) does not seem to be valid for H-plasma in this case. \\Now we consider Barium plasma in a cylinder with $n_0=2\times 10^{13}\,\mathrm{cm^{-3}}$ and $B_0=10^4 \,\mathrm{Gauss}$ for $L_n=10\,\mathrm{cm}$. In this case, $\Omega_i=6.9\times10^5\,\mathrm{rads^{-1}}$, and we choose $R_0=100\,\mathrm{cm}$.\\ The Fig. (5a) shows $\omega$ vs $r$ for Eq. (16) and heavier ions are considered to be stationary because $\Omega_i\ll \omega$ holds. However, the dispersion relation is not valid near $r=0$, the singular point. \\ In Fig. (5b), $\omega$ vs $r$ from Eq. (15) is plotted for Barium plasma taking into account the ion dynamics.\\ Finally, in Fig. (6), we present the plot of Eq. (23) the solution of eigenvalue Eq. (20) for $E_{\theta}$ in a cylinder of $R_0=100 \,\mathrm{cm}$ for JET parameters just as an illustration taking $A_1=1$ and choosing $\omega=3\times 10^9\ \mathrm{s^{-1}}$ from Fig. (3b). This figure shows how the amplitude of rapidly oscillating $E_{\theta}$ with frequency $\omega=3\times 10^9\ \mathrm{s^{-1}}$ decays in the cylindrical (Barium) plasma for $40<r\lesssim R_0$ corresponding to $m=500$. 
\section{Discussion}

The applications of electrostatic and electromagnetic high frequency
drift waves have been discussed with reference to upper F-region of
terrestrial ionosphere and cylindrically bounded laboratory plasmas.
These waves propagate in the direction perpendicular to the external
magnetic field and the density gradient. The waves fall in frequency range of
lower hybrid waves (LHW). The LHW is electrostatic and propagate
obliquely making a small angle with the ambient magnetic field with
$k_{\parallel}\ll k_{\perp}$, but the waves discussed in this work
have $k_{\parallel}=0$.

A great deal of literature exists on waves in inhomogeneous density
plasmas and most of the work has been done on the low frequency
drift waves ($\omega\ll \Omega_i$) both in electrostatic and
electromagnetic limits due to their applications in particle and
energy transport. A few authors have reported two new modes in the
LH frequency range; one is the electrostatic electron convection
mode (Yu et al. 1985) and the other is a purely transverse wave; the magnetic
drift wave (MDW) (Huba 1991; Huba et al. 1995) which has $k_{\parallel}=0$. In a recent paper
(Saleem 2014), it has been presented that the compressible effects should be
added to MDW and then the dispersion relation
will be modified and one finds two different limits of such
electromagnetic drift waves; one is applicable to pure electron
plasma and the other contains the effect of lower hybrid
oscillations.

Here we have pointed out that the electromagnetic waves discussed in
by Saleem (2014) are actually the limiting case of the well-known high
frequency extraordinary (X-) wave. Since X-mode is believed to be a
high frequency short wavelength mode, probably that is the reason
that density gradient effects on this mode are considered to be
small. But it is interesting to look at a low frequency limit of
this mode which gives two important dispersion relations; one is
applicable to purely electron plasmas (Eq.(2) )and the other contains
the ion dynamics (Eq. (6)) as well.

The electrostatic waves have been shown to exist in upper F-region
ionosphere where $1\ll \lambda_e^2 k_y^2$ holds. The elctromagnetic
waves have been applied to a cylindrically bounded plasma with
Gaussian density distribution. For illustration point of view, the
JET plasma parameters have been used for which $\lambda_e^2 k_y^2 < 1$ is satisfied.

The waves discussed here with their applications to both space and
laboratory plasmas can have many more applications. The further
investigation of these waves will be useful for future laboratory
experiments particularly with heavier ion plasmas like barium and
fullerene. They will also be applicable to space plasma
observations.

\pagebreak

\pagebreak
\section*{List of Figures}
\begin{enumerate}
\item Fig.(1a) The plot $\omega$ vs $k_y$ of Eq. (10) shows that $\omega$  is closer to $\Omega_i$ and hence in hydrogen like 
plasmas, the ion dynamics is necessary to be  considered.
\item Fig.(1b) The $\omega$ vs $k_y$ is plotted using Eq. (11) for dynamic ions of H-plasma.
\item Fig.(2a) Plot of Eq. (10) for larger wavelengths still keeping $\lambda_{\omega}\ll L_n$.
\item Fig.(2b) Plot of Eq. (11) for larger wavelengths still keeping $\lambda_{\omega}\ll L_n$.
\item Fig.(3a) The plot for $\omega$ vs $k_y$ with JET parameters show $\Omega_i\lesssim \omega$  and therefore in H-plasma Eq. (2) is not valid.
\item Fig.(3b) For H-plasma with JET parameters $\omega$ vs $k_y$ is plotted with dynamic ions (Eq 6) using Cartesian geometry.
\item Fig.(4a) The plot of pure electron plasma for H-case Eq. (18) as $\omega$ vs $r$ for $ m=500 $ (dotted line) and $ m=2000 $ (solid curve) corresponding to JET parameters. Obviously $\omega$ is closer to $\Omega_i$.
\item Fig.(4b) Here $\omega$ v $r$ is plotted considering H-ions to be dynamic corresponding to JET parameters and we find $\omega<\Omega_i$ . Therefore, Eq.(6) is not valid for H-plasma.
\item Fig.(5a) Eq. (2) is plotted for $\omega$ vs $r$ for Ba plasma and in this case ions can be considered to be stationary because $\Omega_i\ll \omega$ holds.
\item Fig.(5b) The electromagnetic dispersion relation of Eq. (6) is plotted as $\omega$ vs $r$ for Ba plasma considering ions to be dynamic and $\Omega_i<\omega<\Omega_e$ holds.
\item Fig.(6) The radial dependence of rapidly oscillating $E_{\theta}$ of Eq. (23) corresponding to $\omega=3\times 10^9\ {\mathrm{s}}^{-1}$ of Fig. (3b) in cylindrically bounded inhomogeneous plasma.
\end{enumerate}

\end{document}